\documentclass[aps,twocolumn,prb,showpacs,10pt,floatfix,groupedaddress]{revtex4-1}
\usepackage{amssymb}
\usepackage{amsmath}
\usepackage{graphicx}
\usepackage{dcolumn}
\usepackage{bm}
\usepackage{textcomp}
\usepackage{color}

\begin{document}

\title{Generalizing the Fermi velocity of strained graphene from uniform to nonuniform strain}


\author{M. Oliva-Leyva$1$}
\email{moliva@fisica.unam.mx}
\author{Gerardo G. Naumis$1,2$}
\email{naumis@fisica.unam.mx}

\affiliation{1. Departamento de F\'{i}sica-Qu\'{i}mica, Instituto de
F\'{i}sica, Universidad Nacional Aut\'{o}noma de M\'{e}xico (UNAM),
Apartado Postal 20-364, 01000 M\'{e}xico, Distrito Federal,
M\'{e}xico}
\affiliation{2. School of Physics Astronomy and Computational Sciences, George Mason University, Fairfax, Virginia 22030, USA}


\begin{abstract}

The relevance of the strain-induced Dirac point shift to obtain the appropriate
anisotropic Fermi velocity of strained graphene is demonstrated. Then a critical revision 
of the available effective Dirac Hamiltonians is made by studying
in detail the limiting case of a uniform strain. An effective 
Dirac Hamiltonian for nonuniform strain is thus reported, which takes into account 
all strain-induced effects: changes in the
nearest-neighbor hopping parameters, the reciprocal lattice deformation and the true shift 
of the Dirac point. Pseudomagnetic fields are thus explained by means of position-dependent Dirac cones, whereas 
complex gauge fields appear as a consequence of a position-dependent Fermi velocity. Also,
position-dependent Fermi velocity effects on the spinor wavefunction are considered for interesting 
cases of deformations such as flexural modes.

\end{abstract}


\maketitle

\section{Introduction}

Since its discovery,\cite{Novoselov04} graphene has been subject of
many theoretical and experimental studies due to the unique array of
its physical properties.\cite{Geim09,Novoselov11} In particular, the
peculiar relation between its electronic and its mechanical
properties has attracted growing
interest.\cite{Neto09,Vozmediano,Guinea12,Zhan} The unusual long
interval of elastic response\cite{Lee08} makes it possible
observable changes in the electronic structure, such as the opening
of a bandgap\cite{Pereira09a,Colombo} or the merging of Dirac
cones\cite{Montamboux08,Montambaux12}. The strategy is to use strain
engineering as a possibility to guide the electrical transport in
graphene-based devices.\cite{Pereira09b,Lowb,Guinea13,Falko,Nguyen}

In the literature there are different theoretical approaches for
studying the influence of lattice deformations over the
electronic properties of graphene. A quantum field theoretical approach in
curved spaces has been alternatively used to predict electronic
implications due to out-of-plane
deformations.\cite{Cortijo,Vozmediano08,FJ12} Also, methods based
solely on symmetry considerations have been applied to several
problems of strained graphene.\cite{Manes07,Winkler,Linnik,Manes13}
In particular, using group theory techniques, a symmetry analysis
has been performed to construct all the possible terms in the
low-energy effective Hamiltonian for graphene in presence of a
nonuniform strain.\cite{Manes13} More recently, a formulation based
on concepts from discrete differential geometry has shown how the
atomistic structure of two-dimensional crystalline membranes
dictates their mechanical, electronic, and chemical
properties.\cite{Salvador13,SalvadorSSC,SalvadorACS,Salvador14}
Some particular analytical results are available for the case of 
uniaxial strain. For periodic strain, a complex fractal
spectrum with gaps, localization transitions and 
topological states are obtained.\cite{Naumis14}

Nevertheless, the most popular theoretical framework for exploring
the concept of strain engineering combines a nearest-neighbor
tight-binding (TB) model and linear elasticity
theory.\cite{Neto09,Vozmediano} As is well known, this TB-elasticity
approach, in the continuum limit, predicts the existence of
strain-induced pseudomagnetic fields. These pseudomagnetic fields
are described by means of a pseudovector potential
$\bm{A}$ which is related to the strain tensor $\bar{\bm{\epsilon}}$
by\cite{Vozmediano}
\begin{equation}\label{VP}
A_{x}=\frac{\beta}{2a}(\bar{\epsilon}_{xx}-\bar{\epsilon}_{yy}), \ \
\ \ \ A_{y}=-\frac{\beta}{2a}(2\bar{\epsilon}_{xy}),
\end{equation}
where $a$ is the unstrained carbon-carbon distance and $\beta$ is
the electron Gr\"{u}neisen parameter. Thus, nonuniform local
deformations of the lattice can be interpreted as a pseudomagnetic
field, given by $\bm{B}=\nabla\times\bm{A}$ (in units of $\hbar/e$) and perpendicular to the
graphene sample.\cite{Guinea10a,Zenan13,Neek12,Sandler,Zenan14} Scanning tunneling 
microscopy studies in graphene nanobubbles have
reported pseudo Landau levels, which are signatures of
strain-induced pseudomagnetic fields.\cite{Levy,Jiong}

A discussion on the pseudomagnetic field theory was
reactivated due to the explicit inclusion of the local lattice
vectors deformation.\cite{Kitt} Initially, this lattice correction
was supposed to produce an extra pseudovector potential
($\bm{K}_{0}$-dependent), but later on it was shown its physical
irrelevance.\cite{KittE,FJ13,Peeters} Particularly, in Ref.~[\cite{SalvadorSSC}] the
absence of the extra pseudovector potential proposed in the theory
was demonstrated in an explicit manner. Also, the consideration of
the actual atomic positions in the TB Hamiltonian resulted in a more
complete analysis on the position-dependent Fermi velocity for
strained graphene.\cite{FJ13} More recently, another correction has
been identified as important in the derivation of the effective
low-energy Hamiltonian for deformed graphene, by pointing out that
the effective Hamiltonian should be expanded around the true Dirac
point and not around the unperturbed one.\cite{HTY,Our13,Volovik_b,Volovik} 

The principal motivation of the present work is to 
determine the implications of the strain-induced Dirac point shift 
in the derivation of the appropriate anisotropic Fermi velocity.
Moreover, we discuss a possible generalization of the effective Dirac
Hamiltonian for nonuniform in-plane deformations. For that end, we lay out our discussion on a
basic principle: \emph{the theory for graphene under nonuniform
strain should describe the particular case of a uniform strain}.

The paper is organized as follows. In Sec. II we discuss the
effective Dirac Hamiltonian for graphene under a uniform in-plane
strain. By comparing with other approaches available in the literature,
we demonstrate the relevance of the expansion around of the true
Dirac point. In Sec. III we report a generalized effective Dirac
Hamiltonian for graphene under a nonuniform in-plane strain, which 
reproduces the case of a uniform strain. In Sec. IV we summarize the results of our work.

\section{Dirac equation for uniformly strained graphene: critical revision}

To illustrate the derivation of the effective Dirac Hamiltonian in
presence of strain, we first consider graphene under a uniform
strain. We use this particular case as a benchmark to identify the
goodness of any effective Dirac Hamiltonian for strained graphene, even
if the deformation is nonuniform.
In the case of a uniform strain, if $\bm{a}$ represents a general
vector in the unstrained graphene lattice, its strained counterpart
is given by the transformation
\begin{equation}\label{TER}
\bm{a}^{\prime}= (\bar{\bm{I}} + \bar{\bm{\epsilon}})\cdot\bm{a},
\end{equation}
where $\bar{\bm{I}}$ is the $2\times2$ identity matrix and
$\bar{\bm{\epsilon}}$ is the position-independent strain tensor. As
an import example, one can quote the deformation of the three
nearest-neighbor vectors. Selecting the $x$ axis along the
graphene zigzag direction, the unstrained nearest-neighbor vectors
are,
\begin{equation}
\bm{\delta}_{1}=\frac{a}{2}(\sqrt{3},1), \ \
\bm{\delta}_{2}=\frac{a}{2}(-\sqrt{3},1),\ \
\bm{\delta}_{3}=a(0,-1),
\end{equation}
whereas the strained nearest-neighbor vectors can be obtained from
$\bm{\delta}_{n}^{\prime}=(\bar{\bm{I}} + \bar{\bm{\epsilon}})\cdot\bm{\delta}_{n}$, see Fig.~1(a).

On the other hand, a uniform strain distorts the reciprocal space as well. From Eq.~(\ref{TER}) follows that if $\bm{b}$
represents a vector of the unstrained reciprocal lattice, its
deformed counterpart results $\bm{b}^{\prime}=(\bar{\bm{I}} + \bar{\bm{\epsilon}})^{-1}\cdot\bm{b}
\simeq(\bar{\bm{I}} - \bar{\bm{\epsilon}})\cdot\bm{b}$ (see Fig.~1(b)).
However, the high-symmetry points of the Brillouin zone are modified differently. For example, the high-symmetry point 
of the unstrained Brillouin zone $\bm{K}_{0}=(\frac{4\pi}{3\sqrt{3}a},0)$ moves to the new position
$\bm{K}=\frac{4\pi}{3\sqrt{3}a}(1-\bar{\epsilon}_{xx}/2-\bar{\epsilon}_{yy}/2,-2\bar{\epsilon}_{xy})$ under a uniform strain.\cite{Pereira09a}

\begin{figure}[h,t]
\includegraphics[width=8cm]{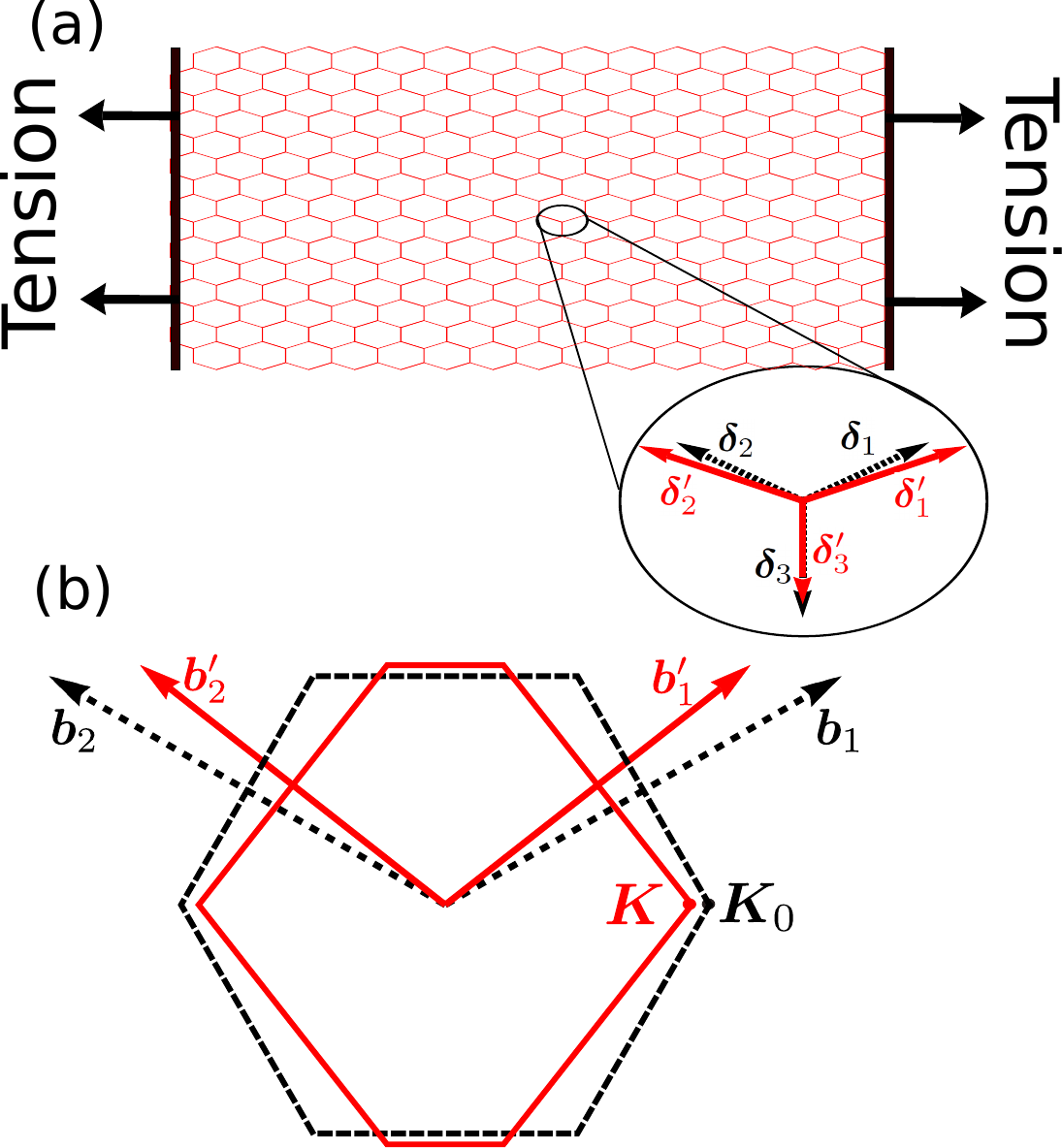}
\caption{\label{fig1} (Color online) (a) Uniaxial stretching along the
zigzag direction of a graphene sample.
The zoom of the honeycomb lattice shows the unstrained $\bm{\delta}_{i}$ (black,
dashed) and strained $\bm{\delta}_{i}^{\prime}$ (red, solid) three
nearest-neighbor vectors. (b) Unstrained (black, dashed) and strained
(red, solid) first Brillouin zone for the same uniaxial zigzag
strain. Note how the reciprocal lattice is contracted along the
direction where the lattice is stretched.}
\end{figure}

For computing the effective Dirac Hamiltonian we start from the
nearest-neighbor TB Hamiltonian,
\begin{equation}
H=-\sum_{\bm{x}^{\prime},n} t_{n}^{\prime}
a_{\bm{x}^{\prime}}^{\dag} b_{\bm{x}^{\prime} +
\bm{\delta}_{n}^{\prime}} + \text{H.c.}, \label{TBH}
\end{equation}
where $\bm{x}^{\prime}$ runs over all sites of the deformed A
sublattice, $a_{\bm{x}^{\prime}}^{\dag}$ is the creation operator
for an electron on the A sublattice at site $\bm{x}^{\prime}$ and
$b_{\bm{x}^{\prime} + \bm{\delta}_{n}^{\prime}}$ is the annihilation operator
for an electron on the B sublattice at site $\bm{x}^{\prime} + \bm{\delta}_{n}^{\prime}$.
The nearest-neighbor hopping parameters $t_{n}^{\prime}$ are modified due to the changes
in intercarbon distance and fulfill an exponential decay,
$t_{n}^{\prime}=t\exp[-\beta(|\bm{\delta}_{n}^{\prime}|/a - 1)]$,
where $t$ is the equilibrium hopping parameter.\cite{Neto09,Ribeiro}

Replacing the creation/annihilation operators with their Fourier
expansions,\cite{Bena} we obtain that the Hamiltonian in momentum
space is given by
\begin{equation}
H=-\sum_{\bm{k},n} t_{n}^{\prime} e^{-i\bm{k}\cdot(\bar{\bm{I}} +
\bar{\bm{\epsilon}})\cdot\bm{\delta}_{n}} a_{\bm{k}}^{\dag}
b_{\bm{k}} + \text{H.c.},\label{k-H}
\end{equation}
and therefore, the closed dispersion relation for uniformly strained
graphene is
\begin{equation}\label{GDR}
E(\bm{k})= \pm\left|\sum_{n} t_{n}^{\prime}
e^{-i\bm{k}\cdot(\bar{\bm{I}} +
\bar{\bm{\epsilon}})\cdot\bm{\delta}_{n}}\right|.
\end{equation}

\begin{figure}[h,t]
\includegraphics[width=8cm]{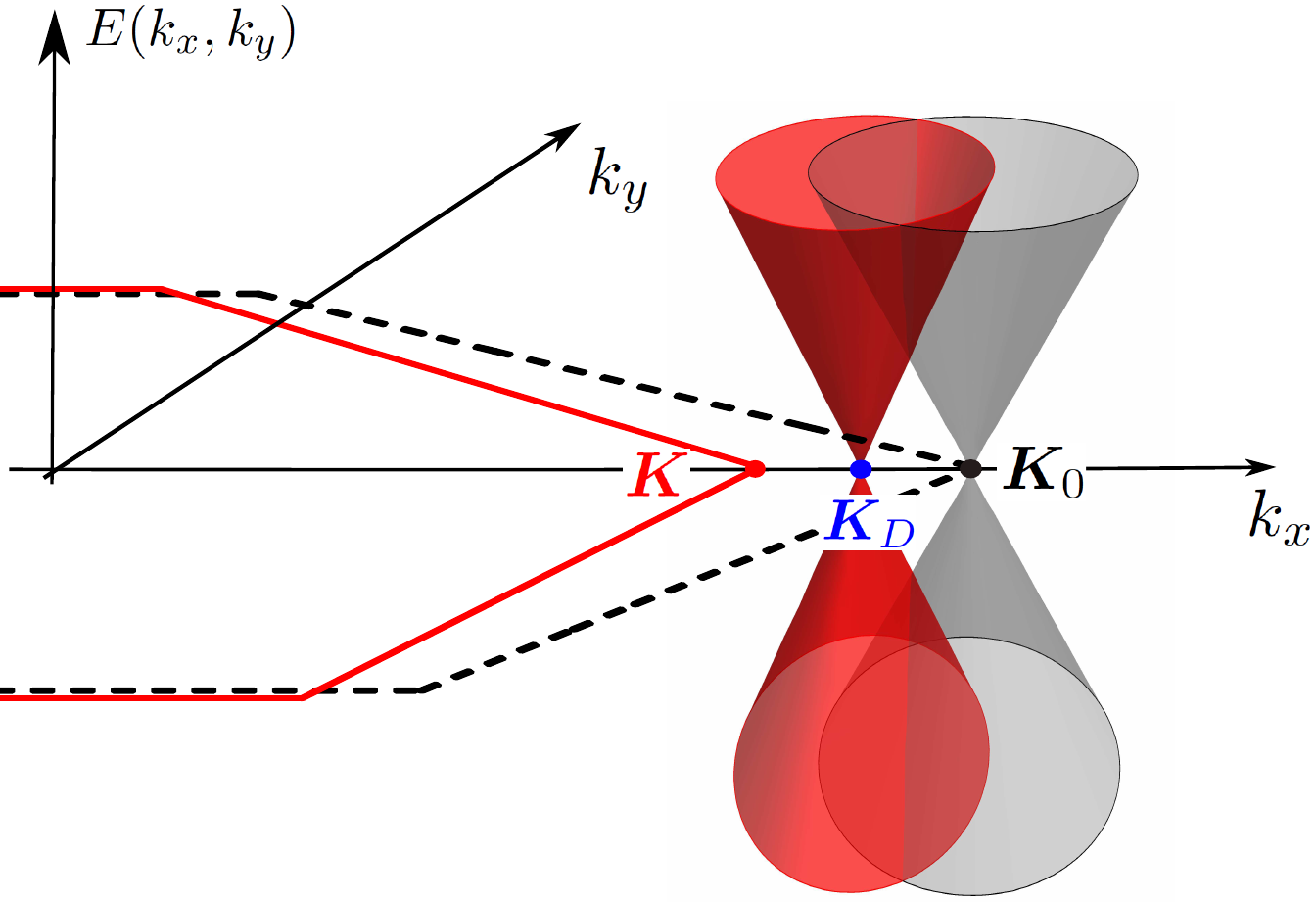}
\caption{\label{fig2} (Color online) Sketch of the Dirac
cone (red cone) movement as  graphene is stretched along the zigzag direction.
Three important points are indicated in the reciprocal space, 
the original Dirac point $\bm{K}_{0}$, the high-symmetry point $\bm{K}$ of
the strained reciprocal lattice and the true Dirac point $\bm{K}_{D}$. The gray Dirac cone is the image of
the red Dirac cone for unstrained graphene. The unstrained reciprocal lattice
is pictured with black dots.}
\end{figure}

As has been documented in other works,\cite{Pereira09a,Our13} the positions
of the minimum of energy, i.e., the $\bm{K}_{D}$ Dirac points
($E(\bm{K}_{D})=0$) obtained from the previous equation, do not coincide with the $\bm{K}$
($\bm{K}_{0}$) high-symmetry points of the strained (unstrained)
Brillouin zone. This is illustrated in Fig. \ref{fig2}. 

Eqs. (\ref{k-H}) and  (\ref{GDR}) are the main ingredients of
the available effective Dirac Hamiltonians. As we will discuss 
below, the main differences come out from the reciprocal-space 
points used for the approximations. This is also illustrated 
in Fig. \ref{fig2}, where the idea is to understand how the
Dirac cone moves and deforms as strain is applied. 

To be more precise, if one considers momenta close to the arbitrary reciprocal-space 
point $\bm{G}$, i.e.
$\bm{k}=\bm{G}+\bm{q}$, the Hamiltonian (\ref{k-H}) can be casted as
\begin{equation}\label{G}
H_{\bm{G}}=-\sum_{n=1}^{3}t_{n}^{\prime}
\begin{pmatrix}
0 & e^{-i(\bm{G}+\bm{q})\cdot(\bar{\bm{I}} +
\bar{\bm{\epsilon}})\cdot\bm{\delta}_{n}}\\
e^{i(\bm{G}+\bm{q})\cdot(\bar{\bm{I}} +
\bar{\bm{\epsilon}})\cdot\bm{\delta}_{n}} & 0
\end{pmatrix},
\end{equation}
and expanding to first order in $\bm{q}$ and
$\bar{\bm{\epsilon}}$, {\it as will be used throughout the rest of the paper}, we obtain
\begin{eqnarray}\label{8a}
H_{\bm{G}}\simeq&-&\sum_{n=1}^{3}t_{n}^{\prime}
\begin{pmatrix}
0 & e^{-i\bm{G}\cdot\bm{\delta}_{n}}\\
e^{i\bm{G}\cdot\bm{\delta}_{n}} & 0
\end{pmatrix}(1+i\sigma_{z}\bm{q}\cdot(\bar{\bm{I}} +
\bar{\bm{\epsilon}})\cdot\bm{\delta}_{n}) \nonumber \\
&\times&
(1+i\sigma_{z}\bm{G}\cdot\bar{\bm{\epsilon}}\cdot\bm{\delta}_{n}),
\end{eqnarray}
with $\sigma_{z}$ being the diagonal Pauli matrix. In the literature,
one can find two kinds of expansions by 
making $\bm{G}=\bm{K}_{0}$ or $\bm{G}=\bm{K}_{D}$.
This leads to two different effective Dirac Hamiltonians, as will be discussed
in the following subsections. 

It is worth mentioning that such Hamiltonians are in fact trying to describe
the deformation and movement of the Dirac cone from different points, 
as explained in Fig.\ref{fig2}.  Clearly, if one chooses a point which is not the 
true Dirac point of the strained system, the Hamiltonian will not display 
the proper symmetries associated with it. Furthermore, {\it one can not pass from
one Hamiltonian to the other by using a simple renormalization of the momentum 
since the Taylor expansions used around each point are different}. 
The Fermi velocity  will be used  to test these ideas.

\subsection{Effective Hamiltonian around $\bm{K}_{0}$}

The most popular expansion in the literature is to consider momenta
close to the high-symmetry points of the unstrained Brillouin zone,
$\bm{G}=\bm{K}_{0}$.\cite{KittE,FJ13,Peeters} In this case,
Hamiltonian (\ref{8a}) can be written as
\begin{eqnarray}\label{8}
H_{\bm{K}_{0}}\simeq&-&\sum_{n=1}^{3}t_{n}^{\prime}
\begin{pmatrix}
0 & e^{-i\bm{K}_{0}\cdot\bm{\delta}_{n}}\\
e^{i\bm{K}_{0}\cdot\bm{\delta}_{n}} & 0
\end{pmatrix}(1+i\sigma_{z}\bm{q}\cdot(\bar{\bm{I}} +
\bar{\bm{\epsilon}})\cdot\bm{\delta}_{n}) \nonumber \\
&\times&
(1+i\sigma_{z}\bm{K}_{0}\cdot\bar{\bm{\epsilon}}\cdot\bm{\delta}_{n}).
\end{eqnarray}

Using the following identity\cite{FJ12,FJ13}
\begin{equation}\label{Iden}
\begin{pmatrix}
0 & e^{-i\bm{K}_{0}\cdot\bm{\delta}_{n}}\\
e^{i\bm{K}_{0}\cdot\bm{\delta}_{n}} & 0
\end{pmatrix}=i\frac{\bm{\sigma}\cdot\bm{\delta}_{n}}{a}\sigma_{z},
\end{equation}
where $\bm{\sigma}=(\sigma_{x},\sigma_{y})$ are the non-diagonal
Pauli matrices, and writing the three nearest-neighbor hopping
parameters $t_{n}^{\prime}$ as
\begin{equation}\label{tn}
t_{n}^{\prime}\simeq
t\left(1-\frac{\beta}{a^2}\bm{\delta}_{n}\cdot\bar{\bm{\epsilon}}\cdot\bm{\delta}_{n}\right),
\end{equation}
the Hamiltonian (\ref{8}) becomes (see Appendix~\ref{AA})
\begin{eqnarray}\label{KF0}
H_{\bm{K}_{0}}&\simeq&\hbar
v_{F}\bm{\sigma}\cdot\left(\bar{\bm{I}}+\bar{\bm{\epsilon}} -
\frac{\beta}{4}(2\bar{\bm{\epsilon}}+\text{Tr}(\bar{\bm{\epsilon}})\bar{\bm{I}})\right)\cdot\bm{q}
- \hbar v_{F}\bm{\sigma}\cdot\bm{A}  \nonumber \\
&+&\hbar
v_{F}\bm{\sigma}\cdot(\frac{a}{2}\bm{K}_{0}\cdot\bar{\bm{\epsilon}}\cdot\bm{\sigma}^{\prime})\cdot\bm{q}
+ \hbar v_{F}\bm{\sigma}\cdot\bar{\bm{\epsilon}}\cdot\bm{K}_{0},
\end{eqnarray}
where $v_{F}=3ta/2\hbar$ is the Fermi velocity for unstrained
graphene and $\bm{\sigma}^{\prime}=(-\sigma_{z},\sigma_{x})$. Here
the $\bm{A}$ vector is given by Eq.~(\ref{VP}) and as mentioned, is
interpreted as a pseudomagnetic vector potential when the strain is
nonuniform. It is worth pointing out that the expression (\ref{VP})
was derived by taking the $x$ axis parallel to the zigzag direction
of the graphene lattice and considering a valley ($\bm{K}_{0}$) with index $+1$. In the 
following, we assume these conditions, unless stated otherwise.

Hamiltonian $H_{\bm{K}_{0}}$ contains a problem that is very easy to spot.
Let us consider a simple isotropic stretching of the lattice, which
can be written as $\bar{\bm{\epsilon}}=\epsilon\bar{\bm{I}}$. This
strain is just a renormalization of the distance between carbons. As
a result, the new carbon-carbon distance under isotropic strain is
$a^{\prime}=a(1+\epsilon)$ and the new hopping parameter to first
order in strain is $t^{\prime}=t(1-\beta\epsilon)$. Thus, the new
Fermi velocity obtained straight away from the nearest-neighbor TB
Hamiltonian is $v_{F}^{\prime}=3t^{\prime} a^{\prime}/2\hbar\simeq
v_{F}(1-\beta\epsilon+\epsilon)$ and therefore, the effective Dirac
Hamiltonian is $\hbar v_{F}^{\prime}\bm{\sigma}\cdot\bm{q}$, which
can not be obtained from Eq.~(\ref{KF0}) to an isotropic strain. This trivial test confirms
that $H_{\bm{K}_{0}}$ is not appropriate to describe graphene under a
uniform strain. Consequently, expansions around the high-symmetry
points of the unstrained Brillouin zone lead to unsuitable effective
Hamiltonians for strained graphene.\cite{Volovik}

\subsection{Effective Dirac Hamiltonian around $\bm{K}_{D}$}

A second option is to derive an effective Dirac
Hamiltonian by expanding (\ref{k-H}) around the true Dirac points.\cite{HTY,Our13,Volovik}
In other words, to make $\bm{G}=\bm{K}_{D}$. As reported in previous work,\cite{Our13}
the actual positions of the $\bm{K}_{D}$
Dirac points to the $\bm{K}_{0}$ point is given by
\begin{equation}\label{NewKD}
\bm{K}_{D}\simeq(\bar{\bm{I}}+\bar{\bm{\epsilon}})^{-1}\cdot\bm{K}_{0} + \bm{A},
\end{equation}
as shown in Fiq.~2. The previous equation confirms the remark that $\bm{K}_{D}$ 
coincides with $\bm{K}$ only for isotropic strain. 

Using Eq.~(\ref{NewKD}) it is possible to obtain the proper
effective Dirac Hamiltonian by developing Eq.~(\ref{k-H}) around the Dirac
points, $\bm{k}=\bm{K}_{D}+\bm{q}$. Following this approach one can
derive that\cite{Our13}
\begin{equation}\label{NewH}
H=\hbar
v_{F}\bm{\sigma}\cdot(\bar{\bm{I}}+\bar{\bm{\epsilon}}-\beta\bar{\bm{\epsilon}})\cdot\bm{q},
\end{equation}
where two strain-induced contributions can be recognized. The
$\beta$-independent term, $\hbar
v_{F}\bm{\sigma}\cdot\bar{\bm{\epsilon}}\cdot\bm{q}$, is purely a
geometric consequence due to lattice deformation and does not depend
of the material as long as it has the same topology. 
On the other hand, the $\beta$-dependent term,
$-\hbar v_{F}\beta\bm{\sigma}\cdot\bar{\bm{\epsilon}}\cdot\bm{q}$,
is owing to the strain-induced changes in the hopping parameters and
its contribution depends of the material since $\beta$ varies
depending on the material. For graphene, both contributions have the 
same order of magnitude.

From Eq.~(\ref{NewH}) one can identify that the appropriate Fermi
velocity tensor is given by
\begin{equation}\label{VecKD}
\bar{\bm{v}}=
v_{F}(\bar{\bm{I}}+\bar{\bm{\epsilon}}-\beta\bar{\bm{\epsilon}}),
\end{equation}
which consistently reproduces the anisotropic transport for
uniformly strained graphene.\cite{Our14,Our14C} For example, Eq.~(\ref{VecKD}) yields
the correct result $v_{F}(1-\beta\epsilon+\epsilon)\bar{\bm{I}}$
when the strain is isotropic,
$\bar{\bm{\epsilon}}=\epsilon\bar{\bm{I}}$. 
Also, for the case of a uniaxial stretching,
\begin{equation}\label{UZ}
\bar{\bm{\epsilon}}=\epsilon
\begin{pmatrix}
1 & 0\\
0 & -\nu
\end{pmatrix},
\end{equation}
with $\nu$ being the Poisson ratio, from Eq.~(\ref{VecKD})
one immediately obtains the known result\cite{Pereira09a,Pereira10}
\begin{equation}\label{VFK}
\bar{\bm{v}}=v_{F}
\begin{pmatrix}
1 + (1-\beta)\epsilon & 0\\
0 & 1-(1-\beta)\epsilon\nu
\end{pmatrix},
\end{equation}
for the anisotropic Fermi velocity. This expression has been used to calculate and explain 
the experimentally observed modulation of the transmittance
of strained graphene with respect to the polarization direction of the incoming
light.\cite{Pereira10,Pereira14,Our15}

\section{Generalized Dirac Hamiltonian for nonuniformly strained graphene}

As mentioned in Sec.~I, we base our discussion for nonuniformly strained graphene
on the following basic principle: \emph{the theory
for graphene under nonuniform strain must describe the particular
case of a uniform strain}. Therefore, one would expect that the
effective Dirac Hamiltonian for nonuniformly strained graphene should reduce to
the effective Dirac Hamiltonian for the case of a spatially uniform strain.
However, none of the effective Dirac Hamiltonian reported in the literature
for the case of a nonuniform strain reduce to Eq.~(\ref{NewH}). This
is an inconsistency in the theory of the strain-induced pseudomagnetic
field, which is owing to expansions around points which are not the true
Dirac points for strained graphene. Below, we give a proposal to 
solve the problem.

Unlike the case of uniform strain, a nonuniform strain
breaks the crystal periodicity. This delicate issue
depends upon the physical considered limit. For example,
if the strain is periodic but with a wavelength comparable to
the interatomic displacement, in certain cases one needs an infinite number
of reciprocal vetors to build the wavefunction, so the present
approach can not be made.\cite{Naumis14} Here we will
assume that the strain modulation wavelength is much bigger than
the interatomic distance, as well as the amplitude.
Under such approximation, the problem is usually
solved by starting from the uniform Hamiltonian and changing $\bar{\bm{\epsilon}}$
to $\bar{\bm{\epsilon}}(\bm{r})$.

The problem lies in the fact that now the Fermi velocity $\bar{\bm{v}}(\bm{r})$ depends 
upon the position, and thus the term $\bar{v}_{ij}q_{k}$ breaks the hermiticity of resulting Hamiltonian. 
To assure hermiticity, the procedure made in previous works to generalize the
Dirac Hamiltonian around the \emph{unstrained Dirac point} $\bm{K_}{0}$,
using the replacement,\cite{FJ12,FJ13,Peeters,HTY}
\begin{equation}
\bar{v}_{ij}q_{k}\rightarrow
\bar{v}_{ij}(\bm{r})\left(-i\frac{\partial}{\partial r_{k}}\right) -
\frac{i}{2}\frac{\partial\bar{v}_{ij}(\bm{r})}{\partial
r_{k}}.
\end{equation}

However, as we discussed previously, the strain-induced
Dirac point shift must be considered in the derivation of the appropriate Fermi velocity.
This issue can be solved by starting from the uniform Hamiltonian around the true Dirac point
in the momentum space, and going to real space by means of the replacement\cite{Volovik}
\begin{equation}\label{rule}
\bar{v}_{ij}q_{k}\rightarrow
\bar{v}_{ij}(\bm{r})\left(-i\frac{\partial}{\partial r_{k}}-K_{k}^{D}(\bm{r})\right) -
\frac{i}{2}\frac{\partial\bar{v}_{ij}(\bm{r})}{\partial
r_{k}},
\end{equation}
where now we have introduced the explicit position-dependence of the Dirac point
by denoting it as $K_{k}^{D}(\bm{r})$. This approach corresponds to the general scheme of emergence 
of gravity and gauge fields in the vicinity of the Weyl, Dirac or Majorana points in the energy spectrum.\cite{Froggatt,Volovik03,Horava}

Thus, according to Eq.~(\ref{rule}) and taking into consideration local rotations
(see Appendix \ref{AC}), the effective Dirac Hamiltonian for
nonuniform in-plane strain can be written as
\begin{equation}\label{HSG}
H=\hbar\bm{\sigma}\cdot\bar{\bm{v}}(\bm{r})\cdot(-i\nabla - \bm{K}_{D}(\bm{r})) - \hbar
v_{F}\bm{\sigma}\cdot\bm{\Gamma},
\end{equation}
where the position-dependent Fermi velocity tensor
$\bar{\bm{v}}(\bm{r})$ is given by
\begin{equation}\label{GV}
\bar{\bm{v}}(\bm{r})=
v_{F}\bigl(\bar{\bm{I}}+\bar{\bm{\epsilon}}(\bm{r})-\beta\bar{\bm{\epsilon}}(\bm{r})\bigr),
\end{equation}
the Dirac point $\bm{K}_{D}(\bm{r})$ by, 
\begin{equation}\label{KDG}
\bm{K}_{D}(\bm{r})=
\bigl(\bar{\bm{I}}-\bar{\bm{\epsilon}}(\bm{r})+\bar{\bm{\omega}}(\bm{r})\bigr)\cdot\bm{K}_{0} + \bm{A}(\bm{r}),
\end{equation}
and the vector field $\bm{\Gamma}$ as
\begin{equation}\label{Gamma}
\Gamma_{i}=\frac{i}{2
v_{F}}\frac{\partial\bar{v}_{ij}(\bm{r})}{\partial
r_{j}}=\frac{i(1-\beta)}{2}\frac{\partial\bar{\epsilon}_{ij}(\bm{r})}{\partial
r_{j}},
\end{equation}
with an implicit sum over repeated indices. 

Let us make some important remarks about our effective Hamiltonian (\ref{HSG}).
First of all, one can see that Eq. (\ref{HSG}) reproduces the limiting case of a uniform strain
in a consistent manner. This is the principal merit of Hamiltonian (\ref{HSG}) with respect to previous effective
Hamiltonians. At the same time, one can recognize a new position-dependent Fermi velocity tensor (Eq. (\ref{GV}))
as the main difference. This is a very important result because enables a more appropriate prediction
of spatially-varying Fermi velocity. Nowadays, such effect of strain has been confirmed by experiments
using scanning tunneling microscopy and spectroscopy.\cite{Hui,Jang}

In the approach carried out, the position-dependent Dirac point generates the pseudomagnetic fields. This fact can be seen
by taking the rotational of the effective potential that appears in Eq.~(\ref{HSG}) which leads to the pseudomagnetic field, 
\begin{equation}
\bm{B}=\nabla\times\bm{K}_{D}(\bm{r}),
\end{equation}
but since $\nabla\times ((\bar{\bm{\epsilon}}(\bm{r})-\bar{\bm{\omega}}(\bm{r}))\cdot\bm{K}_{0})=0$, the 
term  $(\bar{\bm{\epsilon}}(\bm{r})-\bar{\bm{\omega}}(\bm{r}))\cdot\bm{K}_{0}$ does not contribute to the  
pseudomagnetic field. Therefore, the value of the pseudomagnetic field is given by 
\begin{equation}
\bm{B}=\nabla\times\bm{A}(\bm{r}),
\end{equation}
which is exactly the same pseudomagnetic field that appears in other derivations.\cite{FJ12,FJ13,Peeters,HTY} 
Note that, the inclusion of the local rotations tensor $\bar{\bm{\omega}}(\bm{r})$ was necessary to demonstrate the physical
irrelevance of the $\bm{K}_{0}$-dependent pseudovector potential.

On the other hand, the complex gauge field $\bm{\Gamma}$ is owing to a position-dependent Fermi velocity and its
presence guarantees the hermiticity of the Hamiltonian (\ref{HSG}).
Unlike $\bm{A}$, $\bm{\Gamma}$ is a purely imaginary. Thus
$\bm{\Gamma}$ can not be interpreted as a gauge field and will not
give rise to Landau levels in the density of states.\cite{FJ13} 
However, it may have other physical consequences, such as
pseudospin precession, i.e., electronic transitions between the two
sublattices.\cite{FJ13} At present, the experimental signatures
of such complex gauge field $\bm{\Gamma}$ are open questions.

\subsection{Inclusion of out-of-plane deformations}

It is worth mentioning that a second check can be made to Eq. (\ref{HSG}) by adapting an independent
approach developed by Volovik and Zbukov in Ref.[\cite{Volovik}] for out-of-plane deformations. Volovik \emph{et. al.}\cite{Volovik} found a similar Hamiltonian, but 
they used a parametrization thought for curved graphene, where the in-plane coordinates of atoms are identical to their
coordinates in the unperturbed honeycomb lattice. The reason is that they were mainly interested in a
differential geometry interpretation. Here we used the reference laboratory frame, which is more suitable to compare with experiments,
because one must use this frame to describe the interaction with external probes or fields.\cite{FJ13} 
However, once the equations of Volovik and Zubkov are written in the reference laboratory frame,
Eq. (\ref{HSG}) for in-plane deformations can be recovered.

Likewise, one can take advantage of both approaches and to write a generalized effective Dirac Hamiltonian. For this end, in $\beta$-dependent terms
of Eq. (\ref{HSG}) one must replace the strain tensor $\bar{\bm{\epsilon}}$ with the generalized strain tensor 
\begin{eqnarray}
\tilde{\epsilon}_{ij}&=&\frac{1}{2}\left(\frac{\partial u_{i}}{\partial r_{j}} + 
\frac{\partial u_{j}}{\partial r_{i}} + \frac{\partial h}{\partial r_{i}}\frac{\partial h}{\partial r_{j}}\right),\nonumber\\
&=&\bar{\epsilon}_{ij} + \frac{1}{2}\frac{\partial h}{\partial r_{i}}\frac{\partial h}{\partial r_{j}},\label{GTD}
\end{eqnarray}
where $\bm{u}(\bm{r})$ and $h(\bm{r})$ are in- and out-of-plane displacements respectively. 
Thus, finally, the generalized effective Dirac Hamiltonian can be written as
\begin{equation}\label{HG}
H=-i\hbar\bm{\sigma}\cdot\bar{\bm{v}}(\bm{r})\cdot\nabla - \hbar
v_{F}\bm{\sigma}\cdot\bm{A} - \hbar
v_{F}\bm{\sigma}\cdot\bm{\Gamma},
\end{equation}
where now the generalized position-dependent Fermi velocity tensor
$\bar{\bm{v}}(\bm{r})$ results
\begin{equation}\label{VG}
\bar{\bm{v}}(\bm{r})=
v_{F}\bigl(\bar{\bm{I}}+\bar{\bm{\epsilon}}(\bm{r})-\beta\tilde{\bm{\epsilon}}(\bm{r})\bigr),
\end{equation}
with the corresponding complex vector field, 
\begin{equation}
\Gamma_{i}=\frac{i}{2
v_{F}}\frac{\partial\bar{v}_{ij}(\bm{r})}{\partial
r_{j}}=\frac{i}{2}\frac{\partial\bar{\epsilon}_{ij}(\bm{r})}{\partial
r_{j}} -\frac{i\beta}{2}\frac{\partial\tilde{\epsilon}_{ij}(\bm{r})}{\partial
r_{j}}, 
\end{equation}
whereas the pseudovector potential
$\bm{A}$ is given by
\begin{equation}\label{GVP}
A_{x}=\frac{\beta}{2a}(\tilde{\epsilon}_{xx}-\tilde{\epsilon}_{yy}), \ \
\ \ \ A_{y}=-\frac{\beta}{2a}(2\tilde{\epsilon}_{xy}).
\end{equation}

A simple exploration shows that our generalized Hamiltonian (\ref{HG}) reproduces our Hamiltonian for in-plane deformations (Eq. (\ref{HSG})) 
as well as the equations of Volovik and Zubkov,\cite{Volovik} for out-of-plane displacements. 
Note that, we ignored the term $\bigl(\bar{\bm{I}}-\bar{\bm{\epsilon}}(\bm{r})+\bar{\bm{\omega}}(\bm{r})\bigr)\cdot\bm{K}_{0}$ 
due to its demonstrated irrelevance. Consequently, the generalized Hamiltonian (\ref{HG}) describes the particular case of a uniform 
strain which resolves an inconsistency of previous effective Hamiltonians.

\subsection{Effects of position-dependent Fermi velocity on the spinor wavefunction}

Finally, let us now consider the effects of a position-dependent Fermi velocity tensor on the spinor wavefunction of charge carriers. 
For this purpose, we consider the case of a out-of-plane deformation along the $x$ axis given by $h(x)$. Then from Eq. (\ref{GTD}) 
it follows that the generalized strain tensor is,
\begin{equation}
 \tilde{\epsilon}_{xx}(x)= \frac{1}{2}(\partial_{x} h(x))^{2}\equiv f(x)/\beta,\ \ \ \tilde{\epsilon}_{yy} = \tilde{\epsilon}_{xy}=0,
\end{equation}
thus, one immediately obtains that $\bm{A}=(f(x)/(2a),0)$, whereas
\begin{equation}
\bar{\bm{v}}(x)=v_{F}
\begin{pmatrix}
1 -f(x) & 0\\
0 & 1
\end{pmatrix}, \ \ \ \bm{\Gamma}=(-i f'(x)/2,0).
\end{equation}

Taking into consideration that the resulting pseudomagnetic field is zero ($B=\partial_{x}A_{y}-\partial_{y}A_{x}$), from
Eq. (\ref{HG}) one can write the corresponding stationary Dirac equation for the spinor wavefunction $\Psi$ as
\begin{eqnarray}
 \bigl(-i(1- f(x))\partial_{x} - \partial_{y} + i f'(x)/2\bigr)\psi_{2}&=&\varepsilon\psi_{1},\nonumber\\
 \bigl(-i(1- f(x))\partial_{x} + \partial_{y} + i f'(x)/2\bigr)\psi_{1}&=&\varepsilon\psi_{2},\label{LS}
\end{eqnarray}
where the parameter $\varepsilon$ is defined as $\varepsilon\equiv E/(\hbar v_{F})$, and $E$ is the energy. If now
one supposes that the spinor wavefunction is of the form $\Psi=\exp(ik_{y}y)\Phi(x)$ then the following differential equation
system is obtained,
\begin{eqnarray}
 \bigl((1- f(x))\partial_{x} + k_{y} -  f'(x)/2\bigr)\phi_{2}&=&i\varepsilon\phi_{1},\nonumber\\
 \bigl((1-f(x))\partial_{x} - k_{y} -  f'(x)/2\bigr)\phi_{1}&=&i\varepsilon\phi_{2}.\label{LSR}
\end{eqnarray}

In order to recover the case of flat graphene in the appropriate limit one can cast the following
ansatz:
\begin{equation}\label{ansatz}
 \Phi(x)=\exp\Bigl[\int^{x}\frac{i k_{x} + f'(\tilde{x})/2}{1-f(\tilde{x})} d\tilde{x}\Bigr]
 \begin{pmatrix}
c_{1} \\
c_{2}
\end{pmatrix},
\end{equation}
where $c_{1}$ and $c_{2}$ are constants. Consequently, the differential system (\ref{LSR}) becomes the algebraic system
\begin{eqnarray}
 (ik_{x} + k_{y})c_{2}=i\varepsilon c_{1},\nonumber\\
 (ik_{x} + k_{y})c_{1}=i\varepsilon c_{2},
\end{eqnarray}
which has infinite solutions if $\varepsilon=\pm(k_{x}^{2}+k_{y}^{2})^{1/2}$. Therefore, finally we find that
the stationary Dirac equation (\ref{LS}) has as solution the spinor wavefunction
\begin{equation}\label{sol}
 \Psi(\bm{r})= A \exp\Bigl[i k_{y} y + \int^{x}\frac{i k_{x} + f'(\tilde{x})/2}{1-f(\tilde{x})} d\tilde{x}\Bigr]\begin{pmatrix}
1 \\
s e^{i\theta}
\end{pmatrix},
\end{equation}
where $e^{i\theta}=(k_{x}+ik_{y})/|\varepsilon|$, $A$ is a normalization constant and $s=\pm1$ denotes the conduction band 
and valence bands, respectively. 

A remarkable result follows from our solution (\ref{sol}): 
\begin{equation}\label{mod}
 |\Psi|^{2}\sim(1-f(x))^{-1}
\end{equation}
i.e. a position-dependent Fermi velocity induces a inhomogeneity in the carrier probability density. For example,
in the interesting case of a flexural mode given by $h(x)=h_{0}\cos(G x)$, from Eq. (\ref{mod}) one get 
$|\Psi|^{2}\sim(1-\tilde{h}\sin^{2}(G x))^{-1}$, where $\tilde{h}=\beta h_{0}^{2} G^{2}/2$. So that, the carrier probability density
is minimum at the valleys and at the crests of the flexural mode. To end, let us point out that our findings can be easily extended 
to the case of an in-plane deformation (along the $x$) replacing $\beta$ by $\beta -1$.

\section{Conclusions}

In this work we revisited the effective Dirac Hamiltonian for
graphene under a uniform strain, starting from a
tight-binding description. We simultaneously considered three
fundamental strain-induced contributions: \emph{the changes in the
nearest-neighbor hopping parameters, the reciprocal lattice
deformation and the true shift of the Dirac point}.  In particular,
the Dirac point did not coincide with the high-symmetry points of
the strained reciprocal lattice. A detailed discussion about this
last strain-induced effect demonstrates its relevance to obtain the
appropriate Fermi velocity. Finally, we presented a generalized
effective Dirac Hamiltonian for the case of a nonuniform 
deformations. This new Hamiltonian reproduces the case of uniform strain
in the corresponding limit, which was a missing issue in previous
approaches. Within the approach carried out, the strain-induced
pseudomagnetic fields were obtained owing to the floating character 
of the Dirac point $\bm{K}_{D}(\bm{r})$, whereas complex
gauge fields appeared as a consequence of a position-dependent Fermi velocity.
Our expression (\ref{VG}) for the generalized position-dependent Fermi velocity 
tensor is the main result in this paper. Also, we found closed 
analytical solutions  for the spinor wavefunctions in cases of practical interest 
on which the Fermi velocity depends on the position.

\begin{acknowledgments}
We specially thank M. Zubkov and G. Volovik for pointing out a mistake in a previous version of the
manuscript. We also acknowledge conversations with J. E. Barrios and G. Murgu\'{i}a. This
work was supported by UNAM-DGAPA-PAPIIT, project IN-$102513$. M.O.L
acknowledges support from CONACYT (Mexico). G. Naumis thanks a PASPA schoolarship for a
sabatical leave at the George Mason University, where this work has been completed.
\end{acknowledgments}

\appendix

\section{}\label{AA}

In this section, the details of the calculations to derive the effective
Hamiltonian around $\bm{K}_{0}$ are presented. Substituting
Eqs.~(\ref{Iden}) and (\ref{tn}) into Eq.~(\ref{8}) we get
\begin{eqnarray}
H_{\bm{K}_{0}}\simeq&-&t\sum_{n=1}^{3}
(1-\frac{\beta}{a^2}\bm{\delta}_{n}\cdot\bar{\bm{\epsilon}}\cdot\bm{\delta}_{n})
(i\frac{\bm{\sigma}\cdot\bm{\delta}_{n}}{a}\sigma_{z}) \nonumber \\
&\times&(1+i\sigma_{z}\bm{q}\cdot(\bar{\bm{I}} +
\bar{\bm{\epsilon}})\cdot\bm{\delta}_{n})
(1+i\sigma_{z}\bm{K}_{0}\cdot\bar{\bm{\epsilon}}\cdot\bm{\delta}_{n}),\nonumber
\end{eqnarray}
and expanding to first order in strain, $H_{\bm{K}_{0}}$ can be
written as
\begin{eqnarray}
H_{\bm{K}_{0}}\simeq&-&t\sum_{n=1}^{3}
(i\frac{\bm{\sigma}\cdot\bm{\delta}_{n}}{a}\sigma_{z}) \Bigl(
1+i\sigma_{z}\bm{q}\cdot(\bar{\bm{I}} +
\bar{\bm{\epsilon}})\cdot\bm{\delta}_{n}  \nonumber\\
&-&\frac{\beta}{a^{2}}\bm{\delta}_{n}\cdot\bar{\bm{\epsilon}}\cdot\bm{\delta}_{n}
-\frac{\beta}{a^{2}}\bm{\delta}_{n}\cdot\bar{\bm{\epsilon}}\cdot\bm{\delta}_{n}
(i\sigma_{z}\bm{q}\cdot\bm{\delta}_{n}) \nonumber\\
&+&
i\sigma_{z}\bm{K}_{0}\cdot\bar{\bm{\epsilon}}\cdot\bm{\delta}_{n}
-(\bm{K}_{0}\cdot\bar{\bm{\epsilon}}\cdot\bm{\delta}_{n})(\bm{q}\cdot\bm{\delta}_{n})\Bigr).\label{A2}
\end{eqnarray}

Now we collect the contribution of each term of this expression,
\begin{equation}\label{T1}
-t\sum_{n=1}^{3}
(i\frac{\bm{\sigma}\cdot\bm{\delta}_{n}}{a}\sigma_{z})=0,
\end{equation}
\begin{eqnarray}
&-&t\sum_{n=1}^{3}
(i\frac{\bm{\sigma}\cdot\bm{\delta}_{n}}{a}\sigma_{z})
(i\sigma_{z}\bm{q}\cdot(\bar{\bm{I}} + \bar{\bm{\epsilon}})\cdot\bm{\delta}_{n})\nonumber\\
&=&\hbar v_{F}\bm{\sigma}\cdot(\bar{\bm{I}} +
\bar{\bm{\epsilon}})\cdot\bm{q},\label{T2}
\end{eqnarray}
\begin{equation}\label{T3}
t\sum_{n=1}^{3}
(i\frac{\bm{\sigma}\cdot\bm{\delta}_{n}}{a}\sigma_{z})
(\frac{\beta}{a^{2}}\bm{\delta}_{n}\cdot\bar{\bm{\epsilon}}\cdot\bm{\delta}_{n})
=-\hbar v_{F}\bm{\sigma}\cdot\bm{A},
\end{equation}

\begin{eqnarray}
&t&\sum_{n=1}^{3}
(i\frac{\bm{\sigma}\cdot\bm{\delta}_{n}}{a}\sigma_{z})
(\frac{\beta}{a^{2}}\bm{\delta}_{n}\cdot\bar{\bm{\epsilon}}\cdot\bm{\delta}_{n})(i\sigma_{z}\bm{q}\cdot\bm{\delta}_{n})\nonumber\\
&=&-\hbar
v_{F}\frac{\beta}{4}\bm{\sigma}\cdot(2\bar{\bm{\epsilon}}+\text{Tr}(\bar{\bm{\epsilon}})\bar{\bm{I}})\cdot\bm{q},\label{T4}
\end{eqnarray}

\begin{equation}
-t\sum_{n=1}^{3}
(i\frac{\bm{\sigma}\cdot\bm{\delta}_{n}}{a}\sigma_{z})
(i\sigma_{z}\bm{K}_{0}\cdot\bar{\bm{\epsilon}}\cdot\bm{\delta}_{n})
=\hbar v_{F}\bm{\sigma}\cdot\bar{\bm{\epsilon}}\cdot\bm{K}_{0},
\end{equation}

\begin{eqnarray}
&t&\sum_{n=1}^{3}
(i\frac{\bm{\sigma}\cdot\bm{\delta}_{n}}{a}\sigma_{z})
(\bm{K}_{0}\cdot\bar{\bm{\epsilon}}\cdot\bm{\delta}_{n})(\bm{q}\cdot\bm{\delta}_{n})\nonumber\\
&=&\hbar
v_{F}\bm{\sigma}\cdot(\frac{a}{2}\bm{K}_{0}\cdot\bar{\bm{\epsilon}}\cdot\bm{\sigma}^{\prime})\cdot\bm{q},\label{T6}
\end{eqnarray}
where $\bm{\sigma}^{\prime}=(-\sigma_{z},\sigma_{x})$ and the
$\bm{A}$ vector is given by Eq.~(\ref{VP}) if the $x$ axis is
selected parallel to the zigzag direction of the graphene lattice.
Finally, taking into account the contribution of each term in
Eq.~(\ref{A2}), given by Eqs. (\ref{T1})-(\ref{T6}), the effective
Hamiltonian around $K_{0}$ has the form of our Eq.~(\ref{KF0}).

\section{}\label{AC}

In this section, we include the local rotations in the problem of strained graphene.
Note that, under an atomic
displacement field $\bm{u}(\bm{r})$, the strained nearest-neighbor
vectors are given approximately by\cite{KittE}
\begin{equation}\label{Gdeltan}
\bm{\delta}_{n}^{\prime}\simeq(\bar{\bm{I}} +
\bm{\nabla}\bm{u})\cdot\bm{\delta}_{n},
\end{equation}
where $\bm{\nabla}\bm{u}$ is the displacement gradient tensor:
\begin{eqnarray}
[\bm{\nabla}\bm{u}]_{ij}&=&\frac{\partial u_{i}}{\partial r_{j}}
=\frac{1}{2}\left(\frac{\partial u_{i}}{\partial r_{j}}+\frac{\partial u_{j}}{\partial r_{i}}\right)+
\frac{1}{2}\left(\frac{\partial u_{i}}{\partial r_{j}}-\frac{\partial u_{j}}{\partial r_{i}}\right), \nonumber \\
&=&\bar{\epsilon}_{ij}(\bm{r})+\bar{\omega}_{ij}(\bm{r}),\label{U}
\end{eqnarray}
with $\bar{\bm{\omega}}(\bm{r})$ being the rotation tensor, which is
antisymmetric. A position-dependent rotation tensor
$\bar{\bm{\omega}}(\bm{r})$ describes the local rotations associated
to the displacement field, while if $\bar{\bm{\omega}}$ is
independent on the position, it represents a lattice global
rotation which does not have physical implications.

Unlike the strained nearest-neighbor vectors, the three
nearest-neighbor hopping parameters $t_{n}^{\prime}$ do not
dependent on the $\bar{\bm{\omega}}(\bm{r})$ tensor,
\begin{eqnarray}\label{Gtn}
t_{n}^{\prime}&\simeq&
t\left(1-\frac{\beta}{a^2}\bm{\delta}_{n}\cdot\bm{\nabla}\bm{u}\cdot\bm{\delta}_{n}\right),\nonumber\\
&\simeq&
t\left(1-\frac{\beta}{a^2}\bm{\delta}_{n}\cdot\bar{\bm{\epsilon}}(\bm{r})\cdot\bm{\delta}_{n}\right),
\end{eqnarray}
which is an expected result since the rotations do not affect the
module of the nearest-neighbor vectors. Thus, one should expect that
the $\bar{\bm{\omega}}(\bm{r})$ tensor only appears in
$\beta$-independent terms, i.e., in terms of purely geometric
origin.

For our purpose to include the local rotations, let us start with the Hamiltonian of
strained graphene in $\bm{k}$-momentum space,\cite{Peeters}
\begin{equation}\label{HDirac}
H=-\sum_{n=1}^{3}t_{n}^{\prime}
\begin{pmatrix}
0 & e^{-i\bm{k}\cdot(\bar{\bm{I}}+\bm{\nabla}\bm{u})\cdot\bm{\delta}_{n}}\\
e^{i\bm{k}\cdot(\bar{\bm{I}}+\bm{\nabla}\bm{u})\cdot\bm{\delta}_{n}} & 0
\end{pmatrix},
\end{equation}
where $\bar{\bm{\epsilon}}$ and $\bar{\bm{\omega}}$ are considered position-independent. 
In order to obtain the
effective Dirac Hamiltonian one must consider momentum close to the
Dirac point, $\bm{k}=\bm{K}_{D}+\bm{q}$. In this case $\bm{K}_{D}$ can be casted as
\begin{eqnarray}
\bm{K}_{D}&=&\left[(\bar{\bm{I}}+\bm{\nabla}\bm{u})^{\top}\right]^{-1}\cdot(\bm{K}_{0}
+ \bm{A}), \nonumber\\
&\simeq&(\bar{\bm{I}} - \bar{\bm{\epsilon}} + \bar{\bm{\omega}})\cdot\bm{K}_{0}
+ \bm{A},\label{GKD}
\end{eqnarray}
which is a generalization of Eq.(\ref{NewKD}). Substituting Eq.~{\ref{GKD}} into Eq.~{\ref{HDirac}} and
consistently expanding to first order in strain and $\bm{q}$ results in,
\begin{widetext}
\begin{eqnarray}
H&=&-\sum_{n=1}^{3}t_{n}^{\prime}
\begin{pmatrix}
0 & e^{-i(\bm{K}_{0}\cdot\bm{\delta}_{n} +
\bm{q}\cdot(\bar{\bm{I}}+\bm{\nabla}\bm{u})\cdot\bm{\delta}_{n} +
\bm{A}\cdot\bm{\delta}_{n})}\\
e^{i(\bm{K}_{0}\cdot\bm{\delta}_{n} +
\bm{q}\cdot(\bar{\bm{I}}+\bm{\nabla}\bm{u})\cdot\bm{\delta}_{n} +
\bm{A}\cdot\bm{\delta}_{n})} & 0
\end{pmatrix},\nonumber \\
&=&-\sum_{n=1}^{3}t_{n}^{\prime}
\begin{pmatrix}
0 & e^{-i\bm{K}_{0}\cdot\bm{\delta}_{n}}\\
e^{i\bm{K}_{0}\cdot\bm{\delta}_{n}} & 0
\end{pmatrix}(1+ i\sigma_{z}\bm{q}\cdot(\bar{\bm{I}}+\bm{\nabla}\bm{u})\cdot\bm{\delta}_{n})
(1+i\sigma_{z}\bm{A}\cdot\bm{\delta}_{n}).\label{H5}
\end{eqnarray}
\end{widetext}

Using once again the identity (\ref{Iden}) and replacing
$t_{n}^{\prime}$ with the expression (\ref{Gtn}) the Hamiltonian
(\ref{H5}) becomes
\begin{eqnarray}
H&=&-t\sum_{n=1}^{3}
(i\frac{\bm{\sigma}\cdot\bm{\delta}_{n}}{a}\sigma_{z}) \Bigl(1+
i\sigma_{z}\bm{q}\cdot(\bar{\bm{I}}+\bm{\nabla}\bm{u})\cdot\bm{\delta}_{n} \nonumber\\
&-&\frac{\beta}{a^{2}}\bm{\delta}_{n}\cdot\bar{\bm{\epsilon}}\cdot\bm{\delta}_{n}
(i\sigma_{z}\bm{q}\cdot\bm{\delta}_{n})
-(\bm{A}\cdot\bm{\delta}_{n})(\bm{q}\cdot\bm{\delta}_{n})\nonumber\\
&-&\frac{\beta}{a^{2}}\bm{\delta}_{n}\cdot\bar{\bm{\epsilon}}\cdot\bm{\delta}_{n}
+ i\sigma_{z}\bm{A}\cdot\bm{\delta}_{n}\Bigr).\label{IntH}
\end{eqnarray}

The contribution of each term in the last equation is given by
Eqs.~(\ref{T1}), (\ref{T3}), (\ref{T4}) and
\begin{eqnarray}
&-&t\sum_{n=1}^{3}
(i\frac{\bm{\sigma}\cdot\bm{\delta}_{n}}{a}\sigma_{z})
(i\sigma_{z}\bm{q}\cdot(\bar{\bm{I}}+\bm{\nabla}\bm{u})\cdot\bm{\delta}_{n})\nonumber\\
&=&-t\sum_{n=1}^{3}
(i\frac{\bm{\sigma}\cdot\bm{\delta}_{n}}{a}\sigma_{z})
(i\sigma_{z}\bm{q}^{*}\cdot\bm{\delta}_{n}),\nonumber\\
&=&\hbar v_{F}\bm{\sigma}\cdot\bm{q}^{*},\ \ \text{with} \ \
\bm{q}^{*}
=(\bar{\bm{I}}+\bm{\nabla}\bm{u}^{\top})\cdot\bm{q},\nonumber\\
&=&\hbar
v_{F}\bm{\sigma}\cdot(\bar{\bm{I}}+\bm{\nabla}\bm{u}^{\top})\cdot\bm{q},
\end{eqnarray}

\begin{eqnarray}
&t&\sum_{n=1}^{3}
(i\frac{\bm{\sigma}\cdot\bm{\delta}_{n}}{a}\sigma_{z})
(\bm{A}\cdot\bm{\delta}_{n})(\bm{q}\cdot\bm{\delta}_{n})\nonumber\\
&=&-\hbar
v_{F}\frac{\beta}{4}\bm{\sigma}\cdot(2\bar{\bm{\epsilon}}-\text{Tr}(\bar{\bm{\epsilon}})\bar{\bm{I}})\cdot\bm{q},
\end{eqnarray}

\begin{equation}\label{T5}
-t\sum_{n=1}^{3}
(i\frac{\bm{\sigma}\cdot\bm{\delta}_{n}}{a}\sigma_{z})
(i\sigma_{z}\bm{A}\cdot\bm{\delta}_{n}) =\hbar
v_{F}\bm{\sigma}\cdot\bm{A}.
\end{equation}
where it is worth mentioning that the contributions of the last two
terms in Eq.~(\ref{IntH}), Eqs. (\ref{T3}) and (\ref{T5}), cancel.

After looking the
contributions of each term, Eq.~(\ref{IntH}) can be written as
\begin{equation}\label{AGHSG}
H=\hbar v_{F}\bm{\sigma}\cdot(\bar{\bm{I}}+\bm{\nabla}\bm{u}^{\top}-\beta\bar{\bm{\epsilon}})\cdot\bm{q},
\end{equation}
where $\bm{\nabla}\bm{u}^{\top}=\bar{\bm{\epsilon}} -
\bar{\bm{\omega}}$. Hamiltonian
(\ref{AGHSG}) can be considered as the generalization of
Eq.~(\ref{NewH}).

Now to extend Eq. (\ref{AGHSG}) to the case of a nonuniform strain we assume that  
$\bar{\bm{\epsilon}}(\bm{r})$ and $\bar{\bm{\omega}}(\bm{r})$ are position-dependent
and pass to real space by means of the rule\cite{Volovik} 
\begin{equation}\label{ruleA}
\bar{v}_{ij}q_{k}\rightarrow
\bar{v}_{ij}(\bm{r})\left(-i\frac{\partial}{\partial r_{k}}-K_{k}^{D}(\bm{r})\right) -
\frac{i}{2}\frac{\partial\bar{v}_{ij}(\bm{r})}{\partial
r_{k}}.
\end{equation}

In consequence, we obtain that  the effective Dirac
Hamiltonian for nonuniform in-plane strain is given by
\begin{equation}\label{HSGA}
H=\hbar\bm{\sigma}\cdot\bar{\bm{v}}(\bm{r})\cdot(-i\nabla - \bm{K}_{D}(\bm{r})) - \hbar
v_{F}\bm{\sigma}\cdot\bm{\Gamma},
\end{equation}
where 
\begin{equation}\label{GVA}
\bar{\bm{v}}(\bm{r})=
v_{F}\bigl(\bar{\bm{I}}+\bar{\bm{\epsilon}}(\bm{r})-\bar{\bm{\omega}}(\bm{r})-\beta\bar{\bm{\epsilon}}(\bm{r})\bigr),
\end{equation}
\begin{equation}\label{KDGA}
\bm{K}_{D}(\bm{r})=
\bigl(\bar{\bm{I}}-\bar{\bm{\epsilon}}(\bm{r})+\bar{\bm{\omega}}(\bm{r})\bigr)\cdot\bm{K}_{0} + \bm{A}(\bm{r}),
\end{equation}
and 
\begin{equation}\label{GammaA}
\Gamma_{i}=\frac{i}{2
v_{F}}\frac{\partial\bar{v}_{ij}(\bm{r})}{\partial
r_{j}}=\frac{i(1-\beta)}{2}\frac{\partial\bar{\epsilon}_{ij}(\bm{r})}{\partial
r_{j}}-
\frac{i}{2}\frac{\partial\bar{\omega}_{ij}(\bm{r})}{\partial r_{j}},
\end{equation}
with an implicit sum over repeated indices. Finally, we remove the dependence on $\bar{\bm{\omega}}$
from Eqs.~(\ref{GVA}) and (\ref{GammaA}) carrying out the following local rotation of the pseudospinor
\begin{equation}\label{trans}
\psi\rightarrow\exp(\frac{i}{2}\bar{\omega}_{xy}\sigma_{3})\psi\simeq\psi+\frac{i}{2}\bar{\omega}_{xy}\sigma_{3}\psi,
\end{equation}
and as a consequence, Eq. (\ref{HSGA}) takes the form of our Eq. (\ref{HSG}).

\bibliography{biblioStrainedGraphene}

\end{document}